\documentclass[aps,prd,amsmath,showpacs]{revtex4}
\usepackage{amsmath,epsf}
\newcommand{\fr}[2]{\frac{#1}{#2}}
\newcommand{\Ref}[1]{(\ref{#1})}
\newcommand{\be}{\begin{equation}}
\newcommand{\ee}{\end{equation}}
\newcommand{\bn}{\begin{eqnarray}}
\newcommand{\en}{\end{eqnarray}}
\newcommand{\bd}{\begin{displaymath}}
\newcommand{\ed}{\end{displaymath}}
\newcommand{\bnn}{\begin{eqnarray*}}
\newcommand{\enn}{\end{eqnarray*}}
\newcommand{\bml}{\begin{mathletters}}
\newcommand{\eml}{\end{mathletters}}
\newcommand{\adb}{\allowdisplaybreaks }
\newcommand{\bs}{\begin{subequations}}
\newcommand{\es}{\end{subequations}}
\begin{document}
\title{On the uniform asymptotic expansion of the Legendre functions.}

\author{Nail R. Khusnutdinov}\email{nk@dtp.ksu.ras.ru}
\affiliation{Department of Physics, Kazan State Pedagogical
University, Mezhlauk 1, Kazan, 420021, Russia}

\begin{abstract}
An uniform expansion of the Legendre functions of large indices
are considered by using the WKB approach. We obtain the recurrent
formula for the coefficients of uniform expansion and compare them
with the uniform expansion of the Bessel function.
\end{abstract}
 \maketitle

\section{Introduction}\label{Sec1}
An uniform expansion of special functions is very useful
representation of them which is used in many branches of science.
It is well-known, for example, the Debay uniform expansion of the
Bessel functions \cite{AbrSte}. To obtain the uniform expansion
one usually uses the complicate calculations which exploit an
contour integral representation of function (see for example
\cite{BatErdV2}). In this paper we use the WKB approach to obtain
an uniform expansion for the Legendre functions. Previously, this
question was analyzed by Thorne in Ref.\cite{Tho} by using
different approach and in Ref.\cite{BarKamKar} for particular case
of the Legendre equation. We would like to note that this special
case of calculations plays an important role in the so called
functional methods which are at present the most powerful method
(see \cite{Kir01}).

The organization of this article is as follows. First of all, in
Sec.\ref{Sec2} we reobtain the Debay formulas for the uniform
expansion of Bessel function  by using the WKB approach. In
Sec.\ref{Sec3}, we apply the same method to the Legendre functions
and their derivative. The Appendix contains the list of the first
four coefficients in manifest form.

\section{Uniform expansion of the Bessel functions}\label{Sec2}
In this section we reobtain the well-known \cite{AbrSte} uniform
asymptotic expansion for the Bessel functions of second kind $I_n
(n \lambda)$ and $K_n (n  \lambda)$ for large value of $n $.

These functions obey to the following differential equation
\be
W'' + \fr 1 \lambda W' = n^2 \left(1 + \fr 1{\lambda^2}\right) W,
\label{BesEqu}
\ee
where the prime is the derivative with respect $\lambda$.

Let us represent the solution of above equation as a series over
small value of $1/n $:
\be
W = C e^{n  S_{-1} + S_0}
\sum_{k=0}^\infty n ^{-k} \omega_k \label{W}
\ee
with $\omega_0 =1$. Using this expression in Eq. \Ref{BesEqu} we
obtain the chain of equations
\bs\label{Bes}
\bn
S'_{-1} &=& \varepsilon \sqrt{1 + \fr 1{\lambda^2}},\label{BesS-1}\adb\\
S'_{0}  &=& -\fr 1{2 S'_{-1}}\left\{S''_{-1} + \fr 1 \lambda
S'_{-1}\right\},\label{BesS0}\adb\\
\omega'_{k+1} &=& -\fr\varepsilon 2 \left(\fr
{\lambda\omega'_k}{\sqrt{1+\lambda^2}}\right)' - \fr\varepsilon 8
\fr{\lambda (\lambda^2 - 4 )}{(1+\lambda^2)^{5/2}}\omega_k ,
\en
where $\varepsilon = \pm 1$ and $k = 0,1,\dots$ . With new
variable $t=1/\sqrt{1+\lambda^2}$, the last equation may be
rewritten in more simple form
\be \dot{\omega}_{k+1} =
\fr\varepsilon 2 \left(t^2 (1-t^2)\dot{\omega}_{k}\right)^\cdot +
\fr\varepsilon 8 (1 - 5 t^2)\omega_k ,\label{BesOme}
\ee
\es
where the dot denotes the derivative with respect $t$.

The first integral of the Eqs. \Ref{Bes} has the following form
\bn
S_{-1} &=& \varepsilon (\eta (\lambda) + C_{-1}),\nonumber\adb\\
S_{0}  &=& -\fr 14 \ln (1 + \lambda^2) + C_0,\label{SC}\adb\\
\omega_{k+1} &=& \fr\varepsilon 2 t^2 (1-t^2)\dot{\omega}_{k} +
\fr\varepsilon 8 \int_0^t(1 - 5 t'^2)\omega_k(t') dt'  + C_{k+1},
\nonumber
\en
where
\be
\eta (\lambda) = \sqrt{1 + \lambda^2} + \ln \fr \lambda{1 +
\sqrt{1 + \lambda^2}}. \label{Eta}
\ee

To find the set of constants $C_k,\ k=-1,0,\dots$ we take the
limit $\lambda\to \infty$ in our expressions \Ref{W}, \Ref{SC} and
compare them with well-known asymptotic formulas \cite{AbrSte}
\be\label{IKas}
I_n  (n  \lambda) \approx \fr 1{\sqrt{2\pi n  \lambda}} e^{n
\lambda}(1 + O(\fr 1\lambda ))\ ,\ K_n  (n  \lambda) \approx
\sqrt{\fr\pi{2n \lambda}} e^{-n \lambda}(1 + O(\fr 1\lambda )).
\ee

Because of the next term of expansion is $O(1/\lambda)$ we have to
set $C_k = 0$ for $k\ge 1$. Taking this into account we have the
following expression for uniform expansion in the limit
$\lambda\to\infty$:
\be\label{Was}
W \approx \fr 1{\sqrt{\lambda}} e^{\varepsilon n  \lambda} C
e^{\varepsilon n C_{-1} + C_0}.
\ee

Therefore, $\varepsilon = 1$ corresponds to the uniform expansion
of $I_n  (n  \lambda)$ and $\varepsilon = -1$ to the $K_n  (n
\lambda)$. For coincidence the expression \Ref{Was} with the
asymptotic expansions \Ref{IKas} we have to set $C_{-1} = C_0 = 0$
and $C = 1/\sqrt{2\pi n }$ for $\varepsilon = 1$, and $C =
\sqrt{\pi/2 n }$ for $\varepsilon = -1$.

Therefore, we arrive at the following well-known formulas for
uniform expansion of the Bessel functions
\bn
I_n (n  \lambda) &=& \sqrt{\fr t{2\pi n }} e^{n  \eta (\lambda)}
\sum_{k=0}^\infty n ^{-k} \omega_k(t),\label{UniBess}\adb\\
K_n (n  \lambda) &=& \sqrt{\fr{\pi t}{2n }} e^{-n  \eta (\lambda)}
\sum_{k=0}^\infty (-n )^{-k}\omega_k(t), \nonumber
\en
where
\be\label{Omega}
\omega_{k+1} = \fr 12 t^2 (1-t^2)\dot{\omega}_{k} + \fr 18
\int_0^t(1 - 5 t'^2)\omega_k(t') dt'.
\ee

In order to find formulas for derivative of the Bessel functions
we represent them in the form below
\be
\fr 1n  W' = \widetilde{C} e^{n  \widetilde{S}_{-1} +
\widetilde{S}_0}\sum_{k=0}^\infty n ^{-k} \widetilde{\omega}_k .
\ee
Comparing the derivative of Eq. \Ref{W} with respect $\lambda$
with above formula we obtain
\bn
\widetilde{S}_{-1} &=& S_{-1}, \nonumber \adb\\
\widetilde{S}_{0} &=& S_{0} + \ln (\varepsilon S'_{-1}),\nonumber \adb\\
\widetilde{C} &=& \varepsilon C, \adb\\
\widetilde{\omega}_k &=& \omega_k + \fr\varepsilon 2 t(t^2 - 1)
\omega_{k-1} + \varepsilon t^2 (t^2 - 1) \dot{\omega}_{k-1}.
\nonumber
\en

Therefore, with these expressions we arrive at the well-known
formulas for uniform expansion of the derivative of the Bessel
functions
\bn
\fr 1n  I'_n  (n  \lambda) &=& \fr 1{\sqrt{2\pi n  t}}\fr 1\lambda
e^{n \eta (\lambda)} \sum_{k=0}^\infty n ^{-k} \overline{\omega}_k(t),\adb\\
\fr 1n  K'_n  (n  \lambda) &=& -\sqrt{\fr\pi{2n  t}}\fr 1\lambda
e^{-n \eta (\lambda)} \sum_{k=0}^\infty (-n
)^{-k}\overline{\omega}_k(t),\nonumber
\en
where
\be
\overline{\omega}_k = \omega_k + \fr 12 t(t^2 - 1) \omega_{k-1} +
t^2 (t^2 - 1)\dot{\omega}_{k-1}.
\ee

\section{Uniform expansion of the Legendre functions}\label{Sec3}
In this section we employ the same approach for the Legendre
functions. We consider the following equation \be (1-x^2) \Psi'' -
2x \Psi' -\left( n^2\gamma^2 + \frac{n^2}{1-x^2} + 2\xi\right)\Psi
=0, \label{Main} \ee which has appeared in context of quantum
field theory in curved space-time \cite{KhuBor,KhuBez}. Here $x\in
(-1,1)$, $n, \gamma$ and $\xi$ are real numbers, and the prime is
the derivative with respect $x$. The particular case of this
equation for $\xi = 1/8$ has been considered in Ref.
\cite{BarKamKar}.

The solutions of this equation are the Legendre functions first
and second kind:
\be
P^{n}_\mu[x],\ Q^{n}_\mu[x]
\ee
with index
\be
\mu = -\frac 12 + \frac 12 \sqrt{1 - 8\xi - 4n^2\gamma^2}.
\ee
For $\xi = 1/8$ these functions are called the cone functions
\cite{AbrSte}.

We assume $n > 0$ and consider the following two independent
solutions
\bn
p^n_\mu [x] &=& P^{-n}_\mu [x],\label{pq}\adb\\
q^n_\mu [x] &=& \fr{(-1)^n}2 (Q^n_\mu [x] + Q^n_{- \mu -1} [x])\nonumber\adb\\
&=& -\fr\pi{2 \sin\pi\mu} P^n_\mu [-x]. \nonumber
\en
They are real functions for arbitrary $\mu$ and obey the following
Wronskian condition
\bd
W (p^n_\mu, q^n_\mu) = \fr 1{1-x^2}.
\ed

To obtain the uniform expansion of functions \Ref{pq} for large
number $n$ we represent the solution in the WKB form as below
\be
\Psi = Ce^{n S_{-1}(x) + S_0(x)}\sum_{k=0}^\infty n^{-k} \psi_k(x)
\label{Form}
\ee
with $\psi_0(x) = 1$. We would like to note the difference of the
uniform expansion in form \Ref{Form}, which is over inverse degree
of $n$, with that considered by Thorne in Ref.\cite{Tho}. He
obtained an expansion over inverse degree of $\mu + 1/2 = \sqrt{1
- 8\xi - 4n^2\gamma^2}/2$.

Substituting above expression in Eq. \Ref{Main} we obtain the
chain of equations
\bn
S_{-1}' &=&\varepsilon\sqrt{\fr 1{(1-x^2)^2} + \fr{\gamma^2}{1-x^2}},\nonumber\adb\\
S_0' &=& -\fr 12 \left\{\fr{S_{-1}''}{S_{-1}'} -
\fr{2x}{1-x^2}\right\}, \adb\\
\psi_1' &=& -\fr 1{2S_{-1}'}\left\{S_0'^2 + S_0'' - \fr{2x}{1-x^2}
S_0' - \fr{2\xi}{1-x^2} \right\},\nonumber\adb\\
\psi_{k+1}' &=& -\fr\varepsilon 2
\left\{\fr{(1-x^2)\psi_k'}{\sqrt{1 + \gamma^2 (1-x^2)}} \right\}'
+ \psi_1' \psi_k,\ k\geq 1 , \nonumber
\en
where $\varepsilon = \pm 1$.

The first integral of this chain has the following form
\bn
S_{-1}(x) &=& \varepsilon \left[ \gamma \arctan\frac{\gamma
x}{\sqrt{1 + \gamma^2 (1 - x^2)}}\right.\adb\\
&+&\left. \frac 12 \ln \frac{(1+x)(1 + \gamma^2(1 - x) + \sqrt{1 +
\gamma^2(1 - x^2)})}{(1-x)(1 + \gamma^2(1 + x) + \sqrt{1 +
\gamma^2(1 - x^2)})} + C_{-1}\right] \nonumber\adb\\
S_0(x) &=& -\frac 14 \ln (1 + \gamma^2 (1 - x^2))\nonumber\adb\\
\psi_{k+1}(x) &=& C_{k+1}(\varepsilon) - \frac{\varepsilon}{2}
\frac{1 - x^2}{\sqrt{1 + \gamma^2 (1 - x^2)}} \psi'_k(x)\nonumber\adb\\
&+& \varepsilon \int_0^x \left( -\frac{\gamma^2}{8} \left[
\frac{2-{x'}^2}{ (1 + \gamma^2(1 - {x'}^2))^{3/2}} -
\frac{5{x'}^2}{(1 + \gamma^2(1 - {x'}^2))^{5/2}}\right]\right.
\nonumber\adb\\
&+& \left.\frac{\xi}{(1 + \gamma^2(1 - {x'}^2))^{1/2}}\right)
\psi_k(x') dx' . \nonumber
\en
We have already set the constant $C_0 = 0$. This leads to
redefinition the constant $C$, only.

The formulas look simpler in terms of new variable
\be
v = \frac{x}{\sqrt{1 + \gamma^2(1 - x^2)}}\ ,
\ee
instead of $x$. This quantity obeys to inequality: $|v| \leq |x| <
1$. In terms of this variable we have
\begin{subequations}
\bn
S_{-1}(v) &=& \varepsilon\left\{ - \frac 12 \ln \frac{1-v}{1+v} +
\gamma \arctan \gamma v + C_{-1}\right\}, \adb\\
S_0(v) &=& \frac 14 \ln \frac{1+\gamma^2v^2}{1+\gamma^2}, \adb\\
\psi_{k+1}(v) &=& C_{k+1} - \frac \varepsilon 2
\frac{(1-v^2)(1 + \gamma^2 v^2)}{(1 + \gamma^2)} \dot\psi_k(v) \nonumber \adb\\
&+& \frac{\varepsilon \gamma^2}{8(1 + \gamma^2)} \int_0^v dv'
\left\{ 5v'^2 + \frac{1}{\gamma^2} - 1 + (8\xi -1) \frac{1 +
\gamma^2}{\gamma^2 (1 + \gamma^2 v'^2)} \right\}\psi_k(v').
\en
\end{subequations}
In above formulas the dot denotes the derivative with respect new
variable $v$.

In order to find constants $C_k$ we have to compare our formulas
with exact expressions for the Legendre functions at a fixed
point. For this reason we take the limit $x\to 1$ in our formulas
\be\label{PsiUn}
\Psi \approx C \left(\frac{1 - x}2 \right)^{- \varepsilon n/2}
\exp \left[n\varepsilon \left( C_{-1} - \fr 12\ln (\gamma^2 + 1) +
\gamma \arctan\gamma \right)\right]
\ee
and compare them with well-known expressions \cite{BatErdV1} for
the Legendre functions at point $x = 1$:
\bn
p^n_\mu[x] = P^{-n}_\mu[x] &\approx& \frac 1{n!} \left(\frac{1 -
x}2 \right)^{n/2},\label{PQUn} \adb\\
q^n_\mu [x] = \fr{(-1)^n}2 (Q^n_\mu [x] + Q^n_{- \mu -1} [x])
&\approx& \frac {(n-1)!}2 \left(\frac{1 - x}2 \right)^{-n/2}.
\nonumber
\en

Therefore, from Eqs. \Ref{PsiUn}, \Ref{PQUn} we observe that
$\varepsilon = -1$ corresponds to $p^{n}_\mu[x]$ with $C = 1/n!$,
and $\varepsilon = + 1$ corresponds to $q^{n}_\mu[x]$ with $C =
(n-1)!/2$, and
\be
C_{-1} = \frac 12 \ln (1 + \gamma^2) - \gamma \arctan \gamma
\label{14}
\ee
for both signs of $\varepsilon$. Furthermore, the coefficients
$\psi_k(v)$ must obey the following condition
\be
\psi_k(1) = 0.
\ee

Taking into account above formulas  we arrive at the following
expression for uniform expansion of the Legendre's functions
\begin{subequations}\label{P}
\bn
p^n_\mu[x] &=& \frac 1{n!} \left[\frac{1+\gamma^2 v^2}{1 +
\gamma^2} \right]^{1/4} e^{nS_{-1}(v)} \sum_{k=0}^\infty n^{-k}
\psi_k(v) \adb\\
q^n_\mu [x] &=& \fr{(n-1)!}2 \left[\frac{1+\gamma^2 v^2}{1 +
\gamma^2} \right]^{1/4} e^{-nS_{-1}(v)} \sum_{k=0}^\infty
(-n)^{-k} \psi_k(v),
\en
where
\bn
S_{-1}(v) &=& \frac 12 \ln \frac{1-v}{(1+v)(1+\gamma^2)} - \gamma
\left[ \arctan \gamma v - \arctan \gamma \right], \adb\\
\psi_{k+1}(v) &=& \frac{(1-v^2)(1 + \gamma^2 v^2)}{2(1 +
\gamma^2)} \dot\psi_k(v)\label{Pi}\adb\\
&-& \frac{\gamma^2}{8(1 + \gamma^2)} \int_1^v dv' \left\{ 5v'^2 +
\frac{1}{\gamma^2} - 1 + (8\xi -1) \frac{1 + \gamma^2}{\gamma^2 (1
+ \gamma^2 v'^2)} \right\}\psi_k(v'). \nonumber
\en
\end{subequations}

Taking into account the same procedure as we used above for the
derivative of the Bessel functions we obtain the following
formulas for uniform expansion of the derivative of functions
$p^n_\mu$ and $q^n_\mu$
\begin{subequations}\label{PDer}
\bn
\fr 1n \fr d{dx} p^n_\mu [x] &=& -\fr 1{n!}\left[\fr{1+\gamma^2
v^2}{1 + \gamma^2} \right]^{3/4} \fr{1 + \gamma^2}{ 1
-v^2}e^{nS_{-1}(v)}\sum_{k=0}^\infty n^{-k} \overline{\psi}_k(v)\adb\\
\fr 1n \fr d{dx} q^n_\mu [x] &=& \fr{(n-1)!}2\left[\fr{1+\gamma^2
v^2}{1 + \gamma^2} \right]^{3/4} \fr{1 + \gamma^2}{ 1
-v^2}e^{-nS_{-1}(v)}\sum_{k=0}^\infty (-n)^{-k} \overline{\psi}_k(v)\adb\\
\overline{\psi}_k(v) &=& \psi_k(v) - \frac{\gamma^2 v (1 -
v^2)}{2(1 + \gamma^2)}\psi_{k-1}(v) - \frac{(1 - v^2)(1 + \gamma^2
v^2)}{1 + \gamma^2} \dot{\psi}_{k-1}(v)\label{PiDer}.
\en
\end{subequations}
The first four coefficients $\psi_k$ and $\overline{\psi}_k$ are
listed in Appendix.

From the recurrent formula \Ref{Pi} it is possible to find the
value of the coefficients $\psi(v)$ for $\gamma \to \infty$.
Indeed, comparing Eq. \Ref{Pi} in the limit $v \to 0$ and Eq.
\Ref{Omega} in the limit $t\to 1$ we obtain the following relation
\be
\psi_k(0) = (-1)^{k+1} \omega_k(1).
\ee

Now we represent formulas obtained in slightly different form
which is close to expansion the Bessel functions. We set $x=\cos
\epsilon$ and $\gamma = \lambda/\sin \epsilon$ and use the
asymptotic expansion for gamma function from Ref.\cite{BatErdV2}
\bnn
\ln n! &=& (n + \fr 12) \ln n - n + \fr 12 \ln 2\pi +
\sum_{k=1}^\infty \fr{B_{2k}}{2k (2k-1)}\fr 1{n^{2k-1}},\adb\\
\ln (n-1)! &=& (n - \fr 12) \ln n - n + \fr 12 \ln 2\pi +
\sum_{k=1}^\infty \fr{B_{2k}}{2k (2k-1)}\fr 1{n^{2k-1}},
\enn
where $B_k$ are the Bernoulli numbers.

With these notations one has
\bn
p^{n}_\mu[\cos \epsilon] &=& \sqrt{\frac{t}{2\pi n}} e^{n
\tilde{\eta}} \sum_{k=0}^\infty n^{-k} \psi^+_k(v)\left(\frac{\sin
\epsilon}{ \lambda n}\right)^n, \nonumber\adb\\
q^{n}_\mu[\cos \epsilon] &=& \sqrt{\frac{\pi t}{2 n}} e^{-n
\tilde{\eta}} \sum_{k=0}^\infty (-n)^{-k} \psi^+_k(v)
\left(\frac{\sin \epsilon}{ \lambda n}\right)^{-n}, \label{pqe}\adb\\
\frac 1n \fr{dp^{n}_\mu[x]}{dx}_{|x = \cos \epsilon} &=&
-\sqrt{\frac{1}{2\pi nt}} e^{n \tilde{\eta}} \sum_{k=1}^\infty
n^{-k} \overline{\psi}^+_k(v) \left(\frac{\sin \epsilon}{ \lambda
n}\right)^n \frac 1{\sin^2\epsilon}, \nonumber \adb\\
\frac 1n \fr{dq^{n}_\mu[x]}{dx}\strut_{|x = \cos \epsilon} &=&
\sqrt{\frac{\pi}{2 nt}} e^{-n \tilde{\eta}} \sum_{k=1}^\infty
(-n)^{-k} \overline{\psi}^+_k(v) \left(\frac{\sin \epsilon}{
\lambda n}\right)^n \frac 1{\sin^2\epsilon}, \nonumber
\en
where
\begin{subequations}
\bn
\tilde{\eta} &=& \ln \frac{\lambda}{ \sqrt{1 + \lambda^2} + \cos
\epsilon} - \frac{\lambda}{\sin \epsilon} \left[ \arctan
\frac{\sin \epsilon}{\lambda} - \arctan \frac{ \tan
\epsilon}{\lambda t }
\right] + 1,\label{TilEta} \adb\\
t &=& \fr 1{\sqrt{1 + \lambda^2}},\ v = t \cos \epsilon, \adb\\
\mu &=& -\frac 12 + \frac 12 \sqrt{1 - 8\xi -
\fr{4n^2\lambda^2}{\sin^2 \epsilon}},
\en
and the coefficients $\psi^+_k$ are found from relation
\be
\sum_{k=0}^\infty n^{-k} \psi^+_k(v) = \exp \left( -
\sum_{k=1}^\infty \frac{B_{2k}}{2k(2k-1)n^{2k-1}}\right)
\sum_{k=0}^\infty n^{-k} \psi_k(v)
\ee
\end{subequations}
by comparing the same degree of $n$ in the left and right hand
sides.

The expressions \Ref{pqe} have the form similar to that for the
Bessel functions expansion given by Eq. \Ref{UniBess}.
Furthermore, it is easy to see that in the limit $\epsilon \to 0$
(argument of the Legendre functions tends to unit and lower index
tends to infinity) the uniform expansion obtained is transformed
to the uniform expansion of the Bessel functions below
\bn
\lim_{\epsilon\to 0}\mu^n p^n_\mu [\cos\epsilon] &=& i^n I_n
(n\lambda), \adb\\
\lim_{\epsilon\to 0}\mu^{-n} q^n_\mu [\cos\epsilon] &=& i^{-n} K_n
(n\lambda)\nonumber
\en
as it should be according with well-known formulas \cite{BatErdV1}
\bn
\lim_{z\to \infty} z^n P^{-n}_z\left[\cos \fr xz \right] &=&
J_n(x), \label{limit}\adb\\
\lim_{z\to \infty} z^n Q^{-n}_z\left[\cos \fr xz \right] &=&
-\fr\pi 2 Y_n(x),\nonumber
\en
where $x = in\lambda$ and $z = in\lambda /\epsilon$. In this limit
the function $\tilde{\eta}$ given by Eq. \Ref{TilEta} coincides
with function $\eta$ \Ref{Eta} in the uniform expansion of Bessel
functions:
\be
\lim_{\epsilon\to 0}\widetilde{\eta}  = \ln \frac{\lambda}{\sqrt{1
+ \lambda^2} + 1} + \sqrt{1 + \lambda^2}.
\ee

\begin{figure}
\centerline{\epsfxsize=8truecm\epsfbox{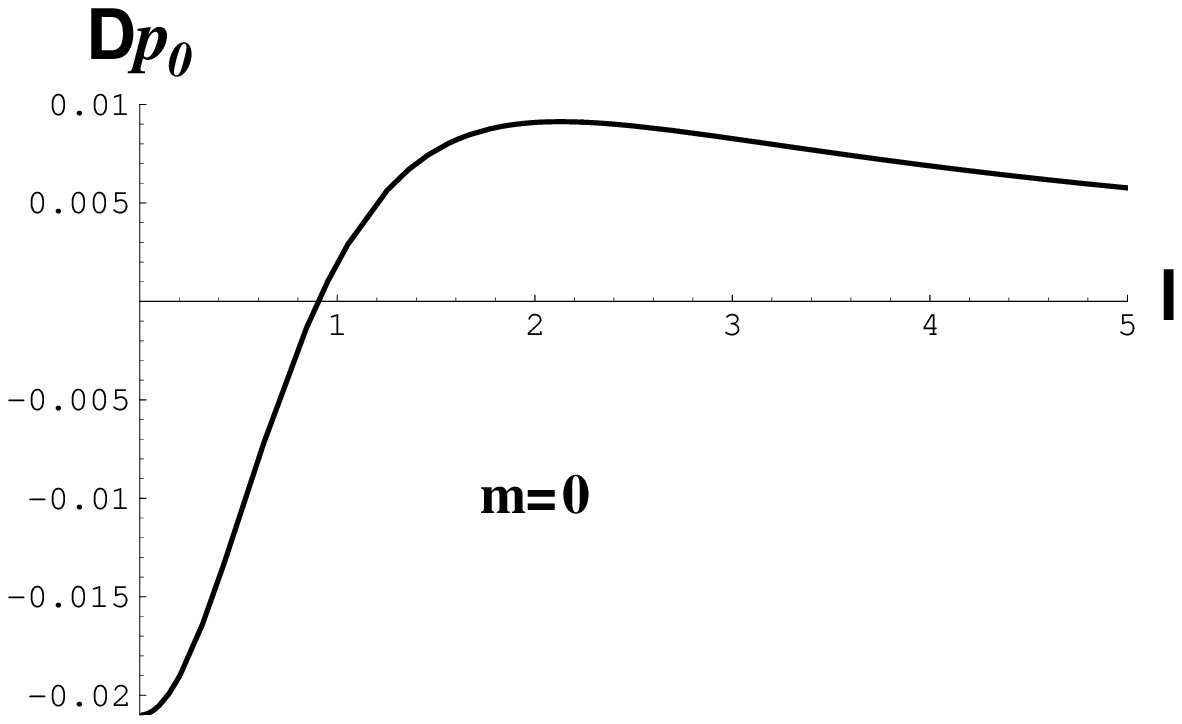}%
\epsfxsize=8truecm\epsfbox{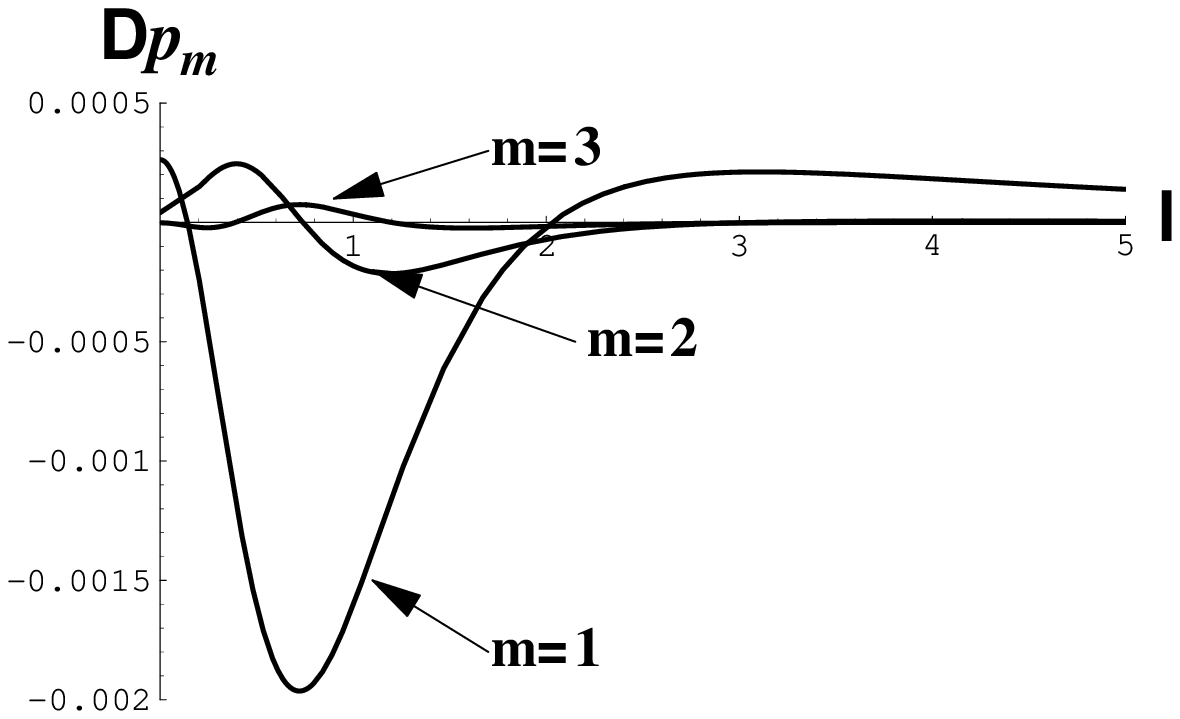}} \caption{The plot of the
relative errors $\Delta p_m = (p^n_\mu - (p^n_\mu)_m )/p^n_\mu$
versus of $\lambda$ for $\epsilon = 0.1,\ \xi = 0$ and $n=4$. Here
$(p^n_\mu)_m$ is the uniform expansion of the Legendre function
$p^n_\mu$ up to degree $m$. }\label{f1}
\end{figure}
\begin{figure}
\centerline{\epsfxsize=8truecm\epsfbox{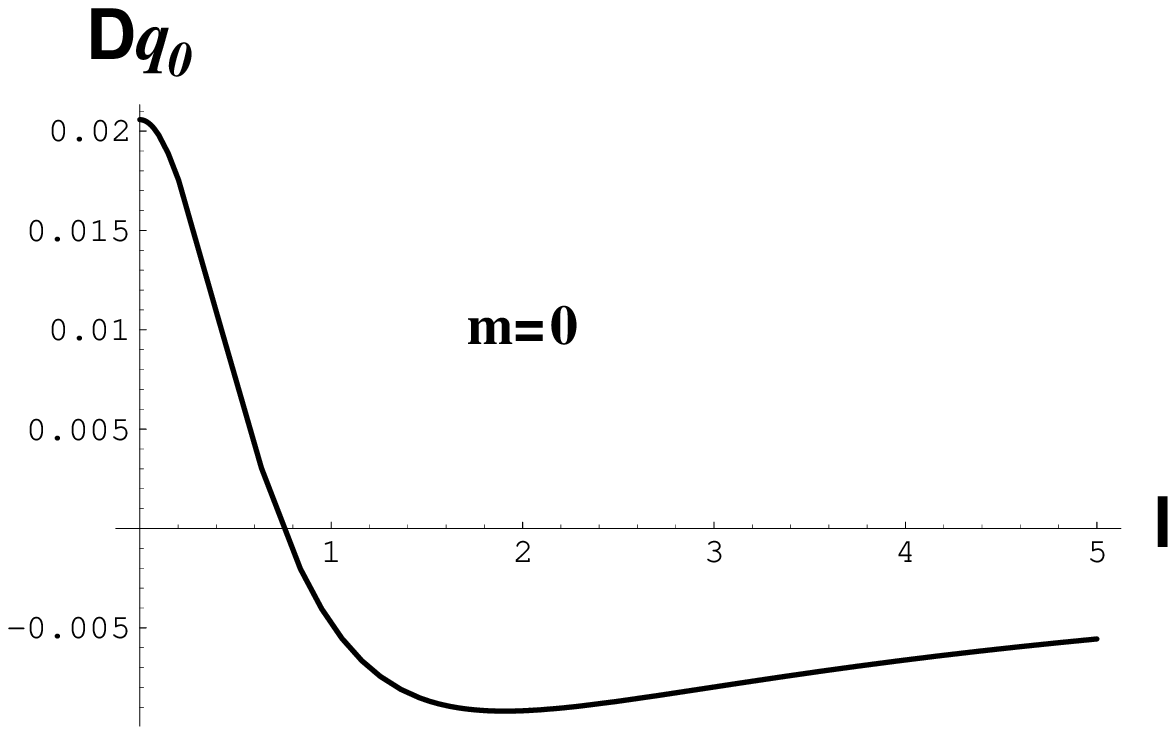}%
\epsfxsize=8truecm\epsfbox{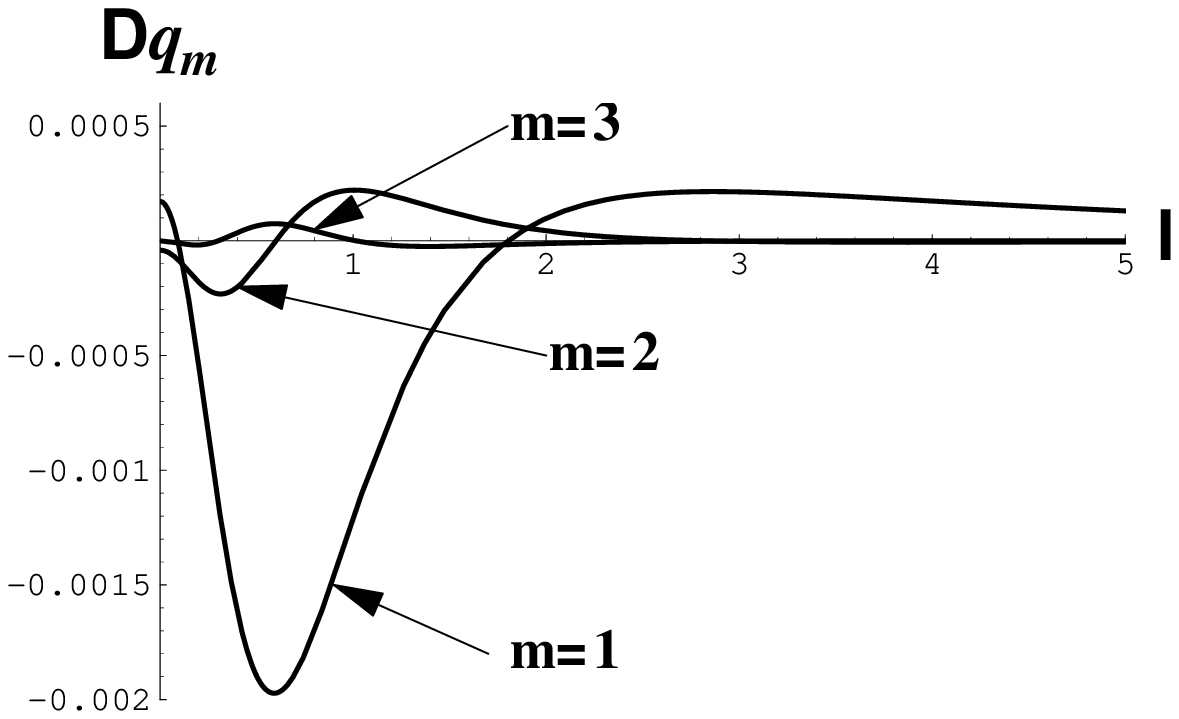}} \caption{The plot of the
relative errors $\Delta q_m = (q^n_\mu - (q^n_\mu)_m) /q^n_\mu$
versus of $\lambda$ for $\epsilon = 0.1,\ \xi = 0$ and $n=4$. Here
$(q^n_\mu)_m$ is the uniform expansion of the Legendre function
$q^n_\mu$ up to degree $m$.}\label{f2}
\end{figure}

The numerical calculation of the relative errors $\Delta p_m =
(p^n_\mu - (p^n_\mu)_m)/p^n_\mu$ and  $\Delta q_m = (q^n_\mu -
(q^n_\mu)_m) /q^n_\mu$ are plotted in Fig. 1 and Fig. 2 for
different $m = 0,1,2,3$ as function $\lambda$, where $(p^n_\mu)_m$
and $(q^n_\mu)_m$ are the uniform expansions of the Legendre
functions $p^n_\mu$ and $q^n_\mu$ up to degree $n^{-m}$. The
difference is smaller the greater $\lambda$.

In conclusion we would like to summarize the results. In this
paper we obtain the uniform expansion for the Legendre functions
$p^n_\mu [x]$ and $q^n_\mu [x]$ given by Eq. \Ref{pq} for large
indices $n$ and $\mu = -\fr 12 + \fr 12 \sqrt{1 - 8\xi -
4n^2\gamma^2}$ as a series over inverse degree on $n$. These
expansions of the functions are given by Eq. \Ref{P} and by Eq.
\Ref{PDer} for their derivatives with respect of argument $x$. The
coefficients of expansion may be found from recurrent chain of
equations \Ref{Pi} and \Ref{PiDer}. The first four coefficients
are listed in Appendix.

\begin{acknowledgements}
The author would like to thank Dr. M. Bordag for stimulation of
this work and for reading this manuscript. The work was supported
by part the Russian Foundation for Basic Research grant N
02-02-17177.
\end{acknowledgements}

\appendix*
\section{Manifest form of first four coefficients.}\label{A}
Below are the expressions for first four coefficients $\psi_k$ and
$\overline{\psi}_k$ in which  we introduced for simplicity the
following notations:
\be
\delta = \arctan [\gamma] - \arctan[\gamma v],\ \zeta = \xi - \fr
18,\ v = \frac{x}{\sqrt{1 + \gamma^2(1 - x^2)}}.
\ee

\bn
\psi_0 &=& 1, \adb\\
\psi_1 &=& \fr{\delta \zeta}{\gamma} + \fr{1}{\gamma^2 + 1}\left[
\fr{2\gamma^2 + 3}{24} + \fr{v(\gamma^2 - 1)}{8} - \fr{5 v^3
\gamma^2 }{24}\right],\nonumber\adb\\
\psi_2 &=& \fr 12\left(\fr{\delta \zeta}{\gamma}\right)^2 +
\fr{\delta \zeta}{\gamma} \fr{1}{\gamma^2 + 1}\left[\fr{2\gamma^2
+ 3}{24} + \fr{v(\gamma^2 - 1)}{8} - \fr{5 v^3 \gamma^2
}{24}\right] +
\fr{\zeta (-1 + v^2)}{2(\gamma^2 + 1)} \nonumber\adb\\
&+&\fr{1}{(\gamma^2 + 1)^2}\left[\fr{4 \gamma^4 + 84 \gamma^2 - 63
}{1152} + \fr{v(\gamma^2 -1)(2 \gamma^2 + 3)}{192} +
\fr{v^2(9\gamma^4 - 58 \gamma^2 + 9)}{128} \right.\nonumber \adb\\
&-&\left. \fr{5v^3 \gamma^2(2 \gamma^2 + 3)}{576} -
\fr{77v^4\gamma^2(\gamma^2 - 1)}{192}+
\fr{385v^6\gamma^4}{1152}\right],\nonumber \adb\\
\psi_3 &=&\fr 16\left(\fr{\delta \zeta}{\gamma}\right)^3 + \fr 12
\left( \fr{\delta \zeta}{\gamma}\right)^2 \fr{1}{\gamma^2 +
1}\left[ \fr{2\gamma^2 + 3}{24} + \fr{v(\gamma^2 - 1)}{8} - \fr{5
v^3 \gamma^2 }{24}\right] + \fr{\delta
\zeta}{\gamma}\left[\fr{1}{(\gamma^2 +
1)^2}\right. \nonumber \adb\\
&\times& \left\{\fr{4 \gamma^4 + 84 \gamma^2 - 63}{1152}+
\fr{v(\gamma^2 -1)(2 \gamma^2 + 3)}{192} + \fr{v^2(9\gamma^4 - 58
\gamma^2 + 9 )}{128} - \fr{5v^3\gamma^2(2 \gamma^2 + 3)}{576}
\right.  \nonumber \adb\\
&-&\left.\left. \fr{77v^4\gamma^2(\gamma^2 - 1)}{192}+
\fr{385v^6\gamma^4}{1152}\right\}+ \fr{\zeta}{\gamma^2 + 1}
\left\{-\fr{2\gamma^2 + 1}{2\gamma^2} + \fr{v^2}{2}\right\}
\right]
+ \fr{\zeta^2 (1-v)}{2\gamma^2(\gamma^2 + 1)} \nonumber \adb\\
&+&\fr{\zeta}{(\gamma^2 + 1)^2}\left[-\fr{2 \gamma^2 + 7}{48} -
\fr{v(3\gamma^2 -11)}{16} + \fr{v^2(2\gamma^2 + 3)}{48} +
\fr{v^3(44
\gamma^2 - 29)}{48}-\fr{35v^5 \gamma^2}{48}\right]\nonumber \adb\\
&+& \fr 1{(\gamma^2 + 1)^3}\left[-\fr{1112 \gamma^6 + 1116
\gamma^4 - 918 \gamma^2 + 5265}{414720} + \fr{v(4\gamma^6 +
728 \gamma^4 - 4323 \gamma^2 + 711)}{9216}\nonumber\right. \adb\\
&+&\fr{v^2 (2\gamma^2 + 3)(9 \gamma^4 - 58 \gamma^2 + 9)}{3072} +
\fr{v^3(2005\gamma^6 - 37671 \gamma^4 + 37566\gamma^2 - 2025)}{27648} \nonumber \adb\\
&-&\fr{77 v^4 (\gamma^2 - 1)(2\gamma^2 + 3)}{4608} - \fr{13 v^5
\gamma^2 (1053 \gamma^4 - 3706\gamma^2 + 1053)}{15360} + \fr{385
v^6 \gamma^4 (2\gamma^2 + 3)}{27648} \nonumber \adb\\
&+&\left.\fr{17017 v^7 \gamma^4 (\gamma^2 - 1)}{9216} - \fr{85085
v^9 \gamma^6}{82944}\right]\nonumber
\en

\bn
\psi^+_0 &=&\psi_0 = 1,\adb\\
\psi^+_1 &=&\psi_1 - \fr 1{12},\nonumber\adb\\
\psi^+_2 &=&\psi_2 - \fr 1{12}\psi_1 + \fr 1{288},\nonumber\adb\\
\psi^+_3 &=&\psi_3 - \fr 1{12}\psi_2 + \fr 1{288}\psi_1 +
\fr{139}{51840}.\nonumber
\en

\bn
\overline{\psi}_0 &=& 1, \adb\\
\overline{\psi}_1 &=& \fr{\delta \zeta}{\gamma} + \fr{1}{\gamma^2
+ 1}\left[ \fr{2\gamma^2 + 3}{24} - \fr{v(3\gamma^2 + 1)}{8} +
\fr{7
v^3\gamma^2 }{24}\right],\nonumber\adb\\
\overline{\psi}_2 &=& \fr 12\left(\fr{\delta
\zeta}{\gamma}\right)^2 + \fr{\delta \zeta}{\gamma}
\fr{1}{\gamma^2 + 1}\left[\fr{2\gamma^2 + 3}{24} - \fr{v(3\gamma^2
+ 1)}{8} + \fr{7 v^3 \gamma^2 }{24}\right]
+ \fr{\zeta (1 - v^2)}{2(\gamma^2 + 1)} \nonumber\adb\\
&+&\fr{1}{(\gamma^2 + 1)^2}\left[\fr{(2\gamma^2 - 27)(2\gamma^2 -
3) }{1152} - \fr{v(3\gamma^2 + 1)(2 \gamma^2 + 3)}{192} -
\fr{v^2(15\gamma^4 - 62 \gamma^2 + 7)}{128} \right.\nonumber \adb\\
&+&\left. \fr{7v^3 \gamma^2(2 \gamma^2 + 3)}{576} +
\fr{v^4\gamma^2(99\gamma^2 - 79)}{192}-
\fr{455v^6\gamma^4}{1152}\right],\nonumber \adb\\
\overline{\psi}_3 &=&\fr 16\left(\fr{\delta
\zeta}{\gamma}\right)^3 + \fr 12 \left( \fr{\delta
\zeta}{\gamma}\right)^2 \fr{1}{\gamma^2 + 1}\left[ \fr{2\gamma^2 +
3}{24} - \fr{v(3\gamma^2 + 1)}{8} + \fr{7 v^3 \gamma^2
}{24}\right] + \fr{\delta\zeta}{\gamma}
\left[\fr{1}{(\gamma^2 + 1)^2}\right. \nonumber \adb\\
&\times& \left\{\fr{(2 \gamma^2 - 3)(2 \gamma^2 - 27)}{1152}-
\fr{v(3\gamma^2 + 1)(2 \gamma^2 + 3)}{192} - \fr{v^2(15\gamma^4 -
62 \gamma^2 + 7 )}{128} + \fr{7v^3\gamma^2(2 \gamma^2 + 3)}{576}
\right.  \nonumber \adb\\
&+&\left.\left. \fr{v^4\gamma^2(99\gamma^2 - 79)}{192} -
\fr{455v^6\gamma^4}{1152}\right\}- \fr{\zeta}{\gamma^2 + 1}
\left\{\fr{1}{2\gamma^2} + \fr{v^2}{2}\right\} \right]
+ \fr{\zeta^2 (1-v)}{2\gamma^2(\gamma^2 + 1)} \nonumber \adb\\
&+&\fr{\zeta}{(\gamma^2 + 1)^2}\left[\fr{2 \gamma^2 -1}{48} +
\fr{v(3\gamma^2 -7)}{16} - \fr{v^2(2\gamma^2 + 3)}{48} -
\fr{v^3(44
\gamma^2 - 25)}{48}+\fr{35v^5 \gamma^2}{48}\right]\nonumber \adb\\
&+& \fr 1{(\gamma^2 + 1)^3}\left[-\fr{1112 \gamma^6 + 5436
\gamma^4 + 1242 \gamma^2 - 1215}{414720} - \fr{v(12\gamma^6 +
904 \gamma^4 - 4281 \gamma^2 + 585)}{9216}\nonumber\right. \adb\\
&-&\fr{v^2 (2\gamma^2 + 3)(15 \gamma^4 - 62 \gamma^2 + 7)}{3072} -
\fr{v^3(2807\gamma^6 - 42897 \gamma^4 + 37458\gamma^2 - 1863)}{27648} \nonumber \adb\\
&+&\fr{v^4 (99\gamma^2 - 79)(2\gamma^2 + 3)}{4608} + \fr{11 v^5
\gamma^2 (1521 \gamma^4 - 4762\gamma^2 + 1241)}{15360} - \fr{455
v^6 \gamma^4 (2\gamma^2 + 3)}{27648} \nonumber \adb\\
&-&\left.\fr{385 v^7 \gamma^4 (51\gamma^2 - 47)}{9216} + \fr{95095
v^9 \gamma^6}{82944}\right]\nonumber
\en

\newpage
\printfigures
\end{document}